\documentclass{elsart}
\usepackage{graphicx}
\newcommand{\tit}[1]{#1,}
\newcommand{\pag}[2]{pp.~#1--#2}
\newcommand{\art}[4]{{\em #1} {\bf #2}, #3--#4} 
\begin{document}      
\sloppy                               
\begin{frontmatter}
\title{Modelling Widely Scattered States in `Synchronized' Traffic Flow and Possible Relevance
for Stock Market Dynamics}
\author[a]{Dirk Helbing,} 
\author[a]{Davide Batic,} 
\author[a]{Martin Sch\"onhof,} 
\author[a]{and Martin Treiber} 
\address[a]{Institute for Economics and Traffic, Dresden University of
Technology, D-01062 Dresden, Germany}
\begin{abstract}
Traffic flow at low densities (free traffic) is characterized by a quasi-one-dimensional
relation between traffic flow and vehicle density, while no such fundamental diagram exists
for `synchronized' congested traffic flow. Instead, a two-dimensional area of widely scattered
flow-density data is observed as a consequence of a complex traffic dynamics.
For an explanation of this phenomenon and transitions between the different traffic phases, 
we propose a new class of molecular-dynamics-like,
microscopic traffic models based on times to collisions and discuss the properties 
by means of analytical arguments. Similar models may help to understand the 
laminar and turbulent phases in
the dynamics of stock markets as well as the transitions among them.
\end{abstract}
\end{frontmatter}

\section{Introduction}

\begin{figure}[htbp]
  \begin{center}
     \includegraphics[width=10cm]{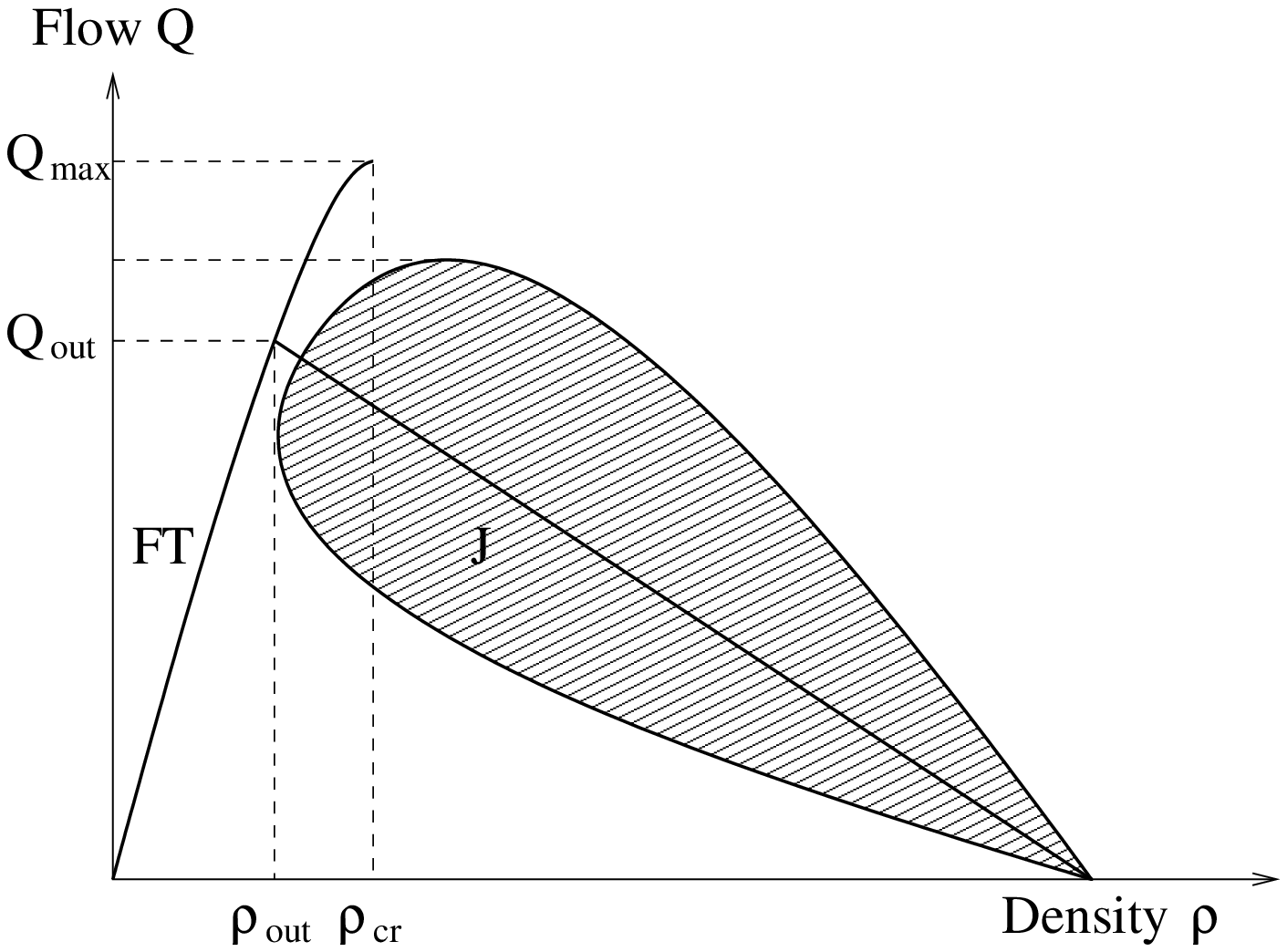}
\end{center}
\caption[]{Schematic illustration of measurements of flow-density data
(one-minute averages) after Kerner.\label{fig1}}
\end{figure}

The dynamics of vehicle traffic has challenged researchers for more than five decades now.
Some major breakthroughs have been made only recently thanks to physicists who applied
methods from statistical physics and non-linear dynamics. In the meantime, 
empirical observations indicate that there are different phases of traffic. According
to Kerner \cite{KerRe96a,KerRe96b,KerRe97,Ker97,Ker98b,Ker99a,KerRe99c,Ker00b}, 
these are (see Fig.~\ref{fig1}): 
\begin{itemize}
\item[1.] free traffic (FT) characterized by a unique flow-density relation $Q_{\rm e}(\rho)$ up to some
maximum flow $Q_{\rm max}$,
\item[2.] wide moving clusters (traffic jams) characterized by a jam line
$J(\rho)$ and `natural constants' such as the
propagation velocity $C< 0$ and the outflow $Q_{\rm out} < Q_{\rm max}$ of wide
clusters, and
\item[3.] `synchronized' congested flow, characterized by a synchronization of average
vehicle speeds among neighboring lanes and a wide scattering of flow-density data. 
\end{itemize}
Synchronized flow has three subtypes: (i) stationary and homogeneous states, (ii) 
oscillatory states, and (iii) homogeneous-in-speed states, where the velocity is
constant, i.e. the flow is proportional
to the density. While the transitions among these subtypes are continuous, the
transitions between free traffic, wide moving jams, and synchronized flow are hysteretic
in nature \cite{Ker97,Ker98b,Ker99a,KerRe99c,Ker00b}. 
\par
Most simulation models show different traffic states as well 
\cite{TGF95,Book,TGF97,TGF99,SchadRev,Review}. Usually, one finds
free and stable traffic at low densities. At higher densities, traffic is unstable in most models, which
can give rise to emergent traffic jams due the slow relaxation to some density-dependent
equilibrium velocity $V_{\rm e}(\rho)$. The delayed adaptation  leads to overbraking
and chain reactions of the following drivers, finally resulting in a standstill of cars. 
In addition to traffic jams, these models can
reproduce congested traffic states of homogeneous and oscillatory type behind bottlenecks
reminding of `synchronized' congested traffic of types (i) and (ii)
\cite{HelHeTr99,TreHeHe00}. The resulting state depends
on the flow on the freeway and on the bottleneck strength, but it is also history-dependent,
as traffic flow can be multi-stable. Despite of this success, some people believe that
the observation of the wide scattering of 
congested flow-density data has not yet been fully understood. Many researchers in the
field of traffic physics have, therefore, published their own theories of `synchronized' flow, 
but none of them is generally accepted,
and the on-going debate on this hot topic is still very controversial.
The proposed explanations include mixtures of
different vehicle types (cars and trucks) \cite{TreHe99a},
a heterogeneity in the time headways \cite{Bank99}, 
changes in the behavior of ``frustrated'' drivers
\cite{Kra98a}, anticipation effects
\cite{Wagner98,Kno00a,Kno00b}, non-unique equilibrium solutions
\cite{Nelson98,Nelson00}, 
and multiple metastable oscillating states \cite{TomSaHa00}.
Only the hypotheses of mixed vehicle types and heterogeneous time headways
have been empirically supported. Here, we will suggest another approach 
to the subject, which is empirically motivated. 
Although we presently do not share some of Kerner's interpretations of 
empirical observations and his criticism of the existing traffic models, 
this paper aims at developing a new class of models
consistent with his interpretation of observed traffic data.

\section{The Times-to-Collision Model}

In the following, we will restrict to identical driver-vehicle units, but
distinguish between two different driver behaviors  corresponding
to different traffic states: 
\begin{itemize}
\item[1.] In free traffic, a vehicle $i$ located at place $x_i(t)$ at time $t$
will try to adapt its actual velocity $v_i = dx_i/dt$
to the so-called optimal velocity $v_{\rm e}(\Delta x_i) = V_{\rm e}(1/\Delta x_i)$ 
within a certain relexation time:
\begin{equation}
  \frac{dv_i}{dt} = \frac{v_{\rm e}(\Delta x_i) - v_i}{\tau} 
\label{free}
\end{equation}
$v_{\rm e}(\Delta x)$ is a function of the headway
(gross/brutto distance) $\Delta x_i = (x_{i-1}-x_i)$ to the
vehicle in front and can be empirically determined. Alternatively, we may use the simple relation
\begin{equation}
  v_{\rm e}(\Delta x) = \max [ v_0,(\Delta x-l)/T ] \, , 
\end{equation}
where $v_0$ is the desired (maximum) velocity in free traffic, $l$ is the
minimum vehicle distance (effective vehicle length) and
$T$ the safe time clearance (net/netto time gap). This corresponds to the idea that a vehicle
tries to keep a velocity-dependent safe distance $(l + T v)$ 
and to drive at maximum speed $v_0$ when
there is enough free space in front.
\item[2.] In `synchronized' congested traffic, vehicles cannot move freely anymore.
Their main concern is to avoid accidents. That is, the time to collision (TTC) given
by $T_i = (\Delta x_i-l)/\Delta v_i$, where $\Delta v_i = (v_i - v_{i-1})$ is the relative velocity,
should not drop below a certain desired value $T_0$, in accordance with observed
driver behavior. In other words: Vehicles tolerate a small (netto) clearance $(\Delta x_i-l)$ to the
vehicle in front, if the relative velocity $\Delta v_i$ is small. Moreover, the headway to the
leading vehicle is history-dependent. It may suddenly shrink due to a lane-changing 
vehicle, 
but usually this does not cause panic braking in order to restore the safe distance
$(l + Tv)$. 
\end{itemize}
Let us first discuss the implications of the empirically justified observation that drivers 
try to maintain a constant
inverse time to collision $1/T_i = \Delta v_i/(\Delta x_i - l)$ during their following behavior.
This implies the equation $d(1/T_i)/dt = 0$, which finally results in
\begin{equation}
  \frac{dv_i}{dt} = \frac{dv_{i-1}}{dt} -  c \frac{\Delta v_i{}^2}{\Delta x_i - l} 
\label{first}
\end{equation}
with $c=\Theta(\Delta v_i)$. The Heaviside step function $\Theta(\Delta v_i)$ is 1 when $\Delta v_i> 0$ and
otherwise 0, thereby reflecting that a vehicle would not brake, when the
relative velocity $\Delta v_i$ is zero or negative (given this is not recommended by
a deceleration $dv_{i-1}/dt < 0$ by the leader). 
\par
We will find that, similar to some other traffic models
\cite{Book,MayKe67,Gip81,KraWaGa97,HelTi98,SauHe98}, 
the $\Delta v_i$-dependent term guarantees safe driving.
Introducing the new variable $z_i = (\Delta x_i - l)$ and 
considering $\Delta v_i = - d\Delta x_i/dt = - dz_i/dt$, we obtain the equation
\begin{equation}
 \frac{d^2 z_i}{dt^2} = - \frac{d\Delta v_i}{dt} = \frac{c}{z_i} \left(\frac{dz_i}{dt}\right)^2
 = - c \, \frac{\Delta v_i}{z_i} \frac{dz_i}{dt}
\label{zeq}
\end{equation}
or
\begin{equation}
 \frac{d \ln (\Delta v_i/v_*)}{dt} = \frac{d\Delta v_i/dt}{\Delta v_i}  = c \frac{dz_i/dt}{z_i} 
 = c \frac{d \ln (z_i/z_*)}{dt} \, ,
\end{equation}
which is analytically solved by 
\begin{equation}
  \frac{\Delta v_i(t)}{\Delta v_i(0)} = \left[ \frac{z_i(t)}{z_i(0)} \right]^c \, .
\end{equation} 
We can immediately see that for any $c>0$, the relative velocity $\Delta v_i$ will 
become zero when  $z_i=0$, i.e. accidents can be avoided for sure. 
For $c=1$, we find 
\begin{equation}
\frac{d\Delta v_i}{dt} = - \frac{\Delta v_i(0)}{\Delta x_i(0) - l}
 \, \Delta v_i \, , 
\end{equation}
while the case of kinematic braking 
with constant deceleration 
\begin{equation}
\frac{d\Delta v_i}{dt} = -b = - \frac{[\Delta v_i(0)]^2}{2[\Delta x_i(0) -l]} < 0
\end{equation} 
corresponds to $c = 1/2$, as usual.
Moreover, {\em the equilibrium solution of equation (\ref{first}) is $\Delta v_i = 0$, but the
distances $\Delta x_i$ between the vehicles in that state are history-dependent and can be very different.
That is, maintaining a certain time to collision
results in driving at some constant velocity as in homogeneous-in-speed states, but
the local density $\rho = 1/\Delta x_i$ may vary considerably.} 
\par
For $c=1$, equation (\ref{zeq}) is solved by 
\begin{equation}
 z_i(t) = z_i(0) \exp \left( - \frac{\Delta v_i(0)}{z_i(0)} t \right) \, ,
\end{equation}
which confirms the history-dependent relaxation behavior. 
If $\Delta v_i > 0$, the headway exponentially
approaches the minimal distance $\Delta x_i = l$. Therefore, we expect platoon formation of
fast cars behind slow ones with $\Delta v_i < 0$. In contrast, the headways in front of slow 
vehicles will usually become so high that they will eventually switch to free driving (see
Sec.~\ref{transi}). Therefore, our model assumes driver reactions 
to the acceleration $dv_{i-1}/dt$ of the leading
vehicle only during following behavior, when the headways are small. As it is hard for
drivers to measure accelerations, it would be reasonable to replace the acceleration of the 
leading vehicle by some step function of $dv_{i-1}/dt$.

\section{Variants of the Model}

We will now introduce an extension of our model of following behavior in `synchronized'
congested traffic. Let us assume that vehicles slowly tend to adapt their time
to collision $T_i = (\Delta x_i-l)/\Delta v_i$ to some preferred value $T_0$.
We may describe this by the equation
\begin{equation}
  \frac{dv_i}{dt} = \frac{dv_{i-1}}{dt} -  c \frac{\Delta v_i{}^2}{\Delta x_i - l} 
 + \omega^2 ( \Delta x_i - l - T_0 \Delta v_i ) 
\label{second}
\end{equation}
with a small ``oscillation frequency'' $\omega$. This equation is equivalent to
\begin{equation}
 \frac{d^2 z_i}{dt^2} + \omega^2 T_0 \frac{dz_i}{dt}
+ \omega^2 z_i = \frac{c}{z_i} \left(\frac{dz_i}{dt}\right)^2 \, ,
\label{back}
\end{equation}
which can be analytically solved. We can distinguish the following cases:
\begin{itemize}
\item[1.] For $c \ne 1$, we can apply the transformation $z_i = y_i^{1/(1-c)}$, which results in
the equation of a damped oscillator:
\begin{equation}
  \frac{d^2y_i}{dt^2} + \omega^2 T_0 \frac{dy_i}{dt} + (1-c) \omega^2 y_i = 0 \, .
\end{equation}
The general solution is, therefore, of the form
\begin{equation}
  z_i(t) = \left( A_1 \mbox{e}^{a_1t} - A_2 \mbox{e}^{-a_2t} \right)^{1/(1-c)} \, ,
\end{equation}
where $A_1 \ge A_2$ are given by the initial conditions and
$a_1 = (\alpha - \omega^2T_0/2)$, $a_2 = (\omega^2T_0/2 + \alpha)$ with
$\alpha^2 = [(\omega^2T_0/2)^2 - \gamma \omega^2]$. When $\alpha^2$ is negative,
the solution describes a damped oscillation, otherwise the relaxation to $z_i = 0$
is overdamped. In the case $\alpha = 0$, the solution is
\begin{equation}
  z_i(t) = B_1 \exp \left( - \frac{\omega^2 T_0}{2\gamma}t \right) (t - B_2)^{1/(1-c)} 
\end{equation}
with parameters $B_1$ and $B_2$ given by the initial conditions.
\item[2.] For $c=1$, the solution reads instead
\begin{equation}
 z_i(t) = C_1 \exp \left[ - \frac{1}{\omega^2T_0}\left(\omega^2 t
+ C_2 \, \mbox{e}^{-\omega^2 T_0 t}\right) \right] \, . 
\end{equation}
The parameters $C_1$ and $C_2 > 0$ are again given by the initial conditions. 
\end{itemize}
In conclusion, we find damped or overdamped non-linear oscillations.
The system oscillates around and/or converges to
$\Delta v_i = 0$ and $z_i = 0$, i.e. $\Delta x_i = l$. Not only would this imply the possibility of
accidents, it also means that the resulting state is characterized by one or several
dense vehicle platoon(s). Instead of this, we would 
prefer oscillations around the safe distance $(l + Tv_i)$ and
$\Delta v_i = 0$. Therefore, we modify equation (\ref{second})
a little: 
\begin{equation}
  \frac{dv_i}{dt} = \frac{dv_{i-1}}{dt} -  c \frac{\Delta v_i{}^2}{\Delta x_i - l}
 + \omega^2 ( \Delta x_i - l - Tv_i - T_0 \Delta v_i ) \, .
\label{third}
\end{equation}
\par
\begin{figure}[htbp]
  \begin{center}
    \includegraphics[width=8cm]{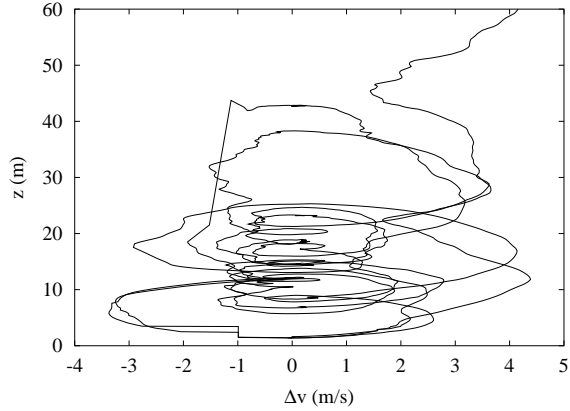}
  \end{center}
    \caption[]{Measured oscillations of the clearance $z = (\Delta x - l)$ and of the relative velocity 
$\Delta v$ around $\Delta v = 0$ (cf. Hoefs, 1972) indicate an instability in
car-following behavior (after Helbing and Tilch, 1998).\label{fig2}}
\end{figure}

Between Eqs. (\ref{second}) and (\ref{third}) lies a major difference. While {\em there is
no stationary solution of equation (\ref{second}) ($\Delta v_i$ always tends to be
non-zero corresponding to a pushy driver behavior),} Eq. (\ref{third}) has the
stationary solution $v_i = (\Delta x_i-l)/T$. A linear stability analysis indicates that, in the
presence of small perturbations, we can expect damped or overdamped oscillations.
Instead, we would prefer an unstable solution leading to permanent oscillations
in the system due to vehicle interactions (see Fig.~\ref{fig2}). In that case we could expect to find a
wide scattering of flow-density data in the congested regime based on a complex
dynamics. 
\par
We can reach this by generalizing equation (\ref{back}) according to
\begin{equation}
 \frac{d^2 z_i}{dt^2} + \mu (z_i)  \frac{dz_i}{dt}
+ \omega^2 ( z_i - Tv_i ) = \frac{c}{z_i} \left(\frac{dz_i}{dt}\right)^2 
\end{equation}
with the a friction function assuming negative values in the neighborhood of
$(\Delta x_i -l) = v_i T$, e.g. 
\begin{equation}
  \mu (z_i) =  \omega^2 T_0 
 + \omega \left[ \beta \left( \frac{Tv_i}{z_i} + \frac{z_i}{Tv_i} \right) - \gamma \right] \, .
\label{friction}
\end{equation}
The term in the square brackets is new and causes a negative friction effect, i.e. a
driving effect, if $\mu < 0$. Such negative friction effects are rather common
in models of traffic and self-driven many-particle systems \cite{Review}. 
By construction, the driving effect can appear in the neighborhood of
$z_i = v_i T$ only if $(\omega T_0 + 2\beta - \gamma) < 0$. 
The larger $(\gamma - 2\beta - \omega T_0)$ is,
the larger will the emerging oscillations be. Because of the scaling of $z_i$ by $v_iT$, they will
also grow with increasing $v_i$, i.e. with local decreasing density $\rho \approx 1/(l+Tv_i)$.
A similar result can be found for related approaches for the friction function
or for $\omega^2$ itself. 
\par
Under the above assumptions, 
the oscillations are limited in size, as the friction coefficient becomes positive, when
the deviation from $z_i = v_iT$ becomes significant. Moreover, the oscillations are
non-linear and influenced by the relative velocity $\Delta v_i = - dz_i/dt$ as well.
The corresponding equation describing the following behavior of driver $i$ reads
\begin{equation}
  \frac{dv_i}{dt} = \frac{dv_{i-1}}{dt} - c \frac{\Delta v_i{}^2 + \eta^2}{\Delta x_i - l}
-  \mu(\Delta x_i-l) \Delta v_i + \omega^2 ( \Delta x_i - l - Tv_i)  
\label{sim}
\end{equation}
with $\eta = 0$. Setting $\eta\ne 0$ allows to
take into account errors in the estimation of $\Delta v_i$ or unexpected variations
of the relative velocity due to lane changes. Moreover,
it would certainly be realistic to add some random noise $\xi_i(t)$ to the acceleration
equations, which will be treated in a forthcoming paper. 
Another reasonable model variant would result by using the acceleration $a = \omega^2
(\Delta x_i - l)$ as a model parameter rather than $\omega$.  
However, the above model shows interesting behavior 
also in the special cases given by $c=1$, $\omega = 0$, $\omega = 1/(\tau T)$,
$\beta = \gamma = 0$, or $\eta = 0$. The other parameters are known 
from empirical investigations.

\section{Transitions Between the Different Traffic States}\label{transi}

It is well-known that the model (\ref{free}) for the driver behavior in free traffic
produces linearly unstable traffic flow and jam formation when the condition
\begin{equation}
  \frac{dv_{\rm e}(\Delta x)}{d\Delta x} > \frac{1}{2\tau}
\end{equation}
is fulfilled. We denote the lowest density $\rho = 1/\Delta x$, 
for which this condition is fulfilled, by
$\rho_{\rm cr}$ (critical density). At this density, the flow $Q_{\rm e}(\rho) = \rho v_{\rm e}(1/\rho)$ 
reaches its maximum $Q_{\rm max}$.  
\par
With some didactically justified simplifications, one can say the following:
Once jams are formed, they are characterized by a jam line 
\begin{equation}
J(\rho) = \frac{1-\rho l}{T}
\end{equation} 
related to $v = v_e(\Delta x) = (\Delta x-l)/T$, 
where $C = -l/T$ corresponds to the backward propagation velocity
of the traffic jam. The lowest density at which the jam line $J(\rho)$ cuts the
function $Q_{\rm e}(\rho)$ defines the characteristic outflow $Q_{\rm out}$ from a traffic jam
and the density $\rho_{\rm out}$ of free flow downstream of it (see Fig.~\ref{fig1}). 
Between the densities
$\rho_{\rm out}$ and $\rho_{\rm cr}$, traffic is metastable, i.e. a perturbation will
grow, if its amplitude exceeds a certain critical amplitude $A(\rho)$, otherwise it will fade away.
The critical amplitude $A(\rho)$ decreases with growing density and becomes zero
at $\rho = \rho_{\rm cr}$. At this density, the probability of a breakdown of traffic is
100\%. 
\par
Transitions from free to `synchronized' congested flow occur more frequently than
transitions from free traffic to wide moving clusters. We will assume that a driver-vehicle
unit switches between free driving and following behavior with probability 1, when
the local flow $Q = v_i/\Delta x_i$ reaches the line 
\begin{equation}
K(\rho) = Q_{\rm max} \left( 1 - \frac{\rho-\rho_{\rm cr}}{1/l - \rho_{\rm cr}} \right) \, , 
\end{equation}
i.e. when the time clearance $(\Delta x_i-l)/v_i$ becomes smaller than 
\begin{equation}
T' = \frac{1-\rho_{\rm cr}l}{Q_{\rm max}}\, . 
\end{equation}
The transition probability shall be zero below
the jam line $J(\rho)$, i.e. when the time clearance is larger than $T$. Between 
the lines $J(\rho)$ and $K(\rho)$, we may assume that the transition probability $P$
is approximately given by
\begin{equation}
  P(\rho, Q) = \left( \frac{Q - J(\rho)}{K(\rho) - J(\rho)}\right)^\epsilon
\end{equation}
with some parameter $\epsilon > 1$. This transition is of hystertic nature.
\par
Finally, we mention that we have checked out deterministic transition rules as well,
e.g. a switching from free to following 
behavior for $\Delta x_i < (l + T' v_i + T'_0\Delta v_i)$
and back to free driving for $\Delta x_i > (l + T v_i +T'_0 \Delta v_i)$, where the
parameter $T'_0$ may agree with $T_0$. The results will be discussed in another paper.

\section{Summary and Outlook}

We have proposed novel microscopic traffic models assuming that the following behavior
in `synchronized' congested flow is different in nature than free driving. The main
assumption is that drivers try to maintain a certain time to collision to guarantee
accident-free driving. The basic model variant had no unique solution, i.e. no fundamental
diagram, and the finally evolving traffic flow was history-dependent. An extended
model variant had a stationary solution, but an unstable one, if the parameters were
appropriately specified. In that case, the vehicle velocities
were non-linearly oscillating with medium-sized amplitudes around the stationary solution. 
The conditions for transitions between free driving and following behavior were specified
in a way inspired by the hypotheses on congested traffic by Kerner 
\cite{Ker97,Ker98b,Ker99a,KerRe99c,Ker00b}. In particular,
we should have hysteretic transitions between free traffic, wide moving clusters, and
`synchronized' flow. Transitions to traffic jams should be rare and require a triggering
by sufficiently large perturbations. 
\par
Based on our present results, all of this appears to be reproducible
with the proposed models. Corresponding simulation results 
will be presented in a forth-coming paper.
Free traffic is characterized by a one-dimensional
flow-density relation of positive slope, wide moving clusters by a
linear jam line of negative slope, and `synchronized' congested flow by
a two-dimensional variation of flow-density data. 
\par
We believe that the transition between laminar regimes (free traffic) and strongly varying regimes
(`synchronized' flow) may have some analogies with the dynamics of stock markets,
where one observes time periods of low and of high volatility 
\cite{GhasBrPeTa96,LuxMa99}. The models suggested above
may, therefore, also help to gain a better understanding of stock markets.

\section*{Acknowledgments}

The authors would like to thank Boris Kerner 
and Andrzej Krawiecki
for inspiring discussions, and Torsten Werner for preparing illustration 1. They are also grateful  
for financial support by the DFG (He 2789/2-1) 
and floating car data by the Bosch GmbH.

\end{document}